# Orbital-Angular-Momentum Embedded Massive MIMO: Achieving Multiplicative Spectrum-Efficiency for mmWave Communications

Wenchi Cheng, *Member, IEEE*, Hailin Zhang, *Member, IEEE*, Liping Liang, Haiyue Jing, and Zan Li, *Senior Member, IEEE*

*Abstract*—By enabling very high bandwidth for radio communications, the millimeter-wave (mmWave), which can easily be integrated with massive-multiple-input-multiple-output (massive-MIMO) due to small antenna size, has been attracting growing attention as a candidate for the fifth-generation (5G) and 5G-beyond wireless communications networks. On the other hand, the communication over the orthogonal states/modes of orbital angular momentum (OAM) is a subset of the solutions offered by massive-MIMO communications. Traditional massive-MIMO based mmWave communications didn't concern the potential spectrum-efficiency-gain (SE-gain) offered by orthogonal states of OAM. However, the highly expecting maximum SE-gain for OAM and massive-MIMO communications is the product of SE-gains offered by OAM and multiplexing-MIMO. In this paper, we propose the OAM-embedded-MIMO (OEM) communication framework to obtain the multiplicative SE-gain for joint OAM and massive-MIMO based mmWave wireless communications. We design the parabolic antenna for each uniform circular array antenna to converge OAM signals. Then, we develop the mode-decomposition and multiplexing-detection scheme to obtain the transmit signal on each OAM-mode of each transmit antenna. Also, we develop the OEM-water-filling power allocation policy to achieve the maximum multiplicative SE-gain for OEM communications. The extensive simulations obtained validate and evaluate our developed parabolic antenna based converging method, mode-decomposition and multiplexing-detection scheme, and OEM-water-filling policy, showing that our proposed OEM mmWave communications can significantly increase the spectrum-efficiency as compared with traditional massive-MIMO based mmWave communications.

## I. INTRODUCTION

THE millimeter-wave (mmWave) with frequencies between 30 and 300 GHz has been attracting growing attention as a candidate for the fifth-generation (5G) and 5G-beyond wireless communications networks [1]. Due to the high available bandwidth and small wavelength, the mmWave can potentially provide high capacity as compared with the traditional radio wireless communications [2]. The massive-multiple-input-multiple-output (massive-MIMO) can be easily implemented in mmWave communications because the size of mmWave antenna can be made to be very small [3], [4].

However, although the mmWave can increase the capacity, it resorts to high bandwidth which cannot increase the spectrum-efficiency (SE) for wireless communications. In fact, as the traditional plane-electromagnetic (PE) wave based wireless communications are becoming more and more mature, it is now very hard to significantly increase the SE of traditional PE wave (which has linear momentum) based wireless communications to meet the rapidly increasing of SE required by the extremely high-data traffics and tremendous amount of users [5]. The released International Mobile Telecommunications 2020 (IMT-2020) shows that the expected SE-gain for the 5G wireless networks is only 3-fold as compared with the fourth-generation (4G) wireless networks [6].

Fortunately, the electromagnetic wave not only has the characteristic of linear momentum which has been studied over a century, but also possesses the angular momentum which attracted much attention during the past decade [7]–[12]. The orbital angular momentum (OAM), which is as a result of a signal possessing helical phase fronts and has not been well studied yet, is another important property of electromagnetic wave [7]–[9]. The OAM-based vorticose wave has different topological charges, which are independent and orthogonal to each other, bridging a new way to significantly increase the SE of wireless communications using different OAM-modes [12]–[18].

Although the OAM-based vorticose radio transmission can increase the SE of wireless communications, some academic researchers have shown that the communication over the orthogonal states of OAM is a subset of the solutions offered by multiple-input-multiple-output (MIMO) communications, thus offering no SE-gain as compared with the traditional multiplexing-MIMO communications [10], [11], [19]. However, noticing that the array-elements of one OAM-generation plate are fed with the same input signal [10], the OAM signal can be generated within one antenna which has several array-elements but only one RF-chain [20]. Thus, the OAM communications can be implemented within one array-elements-based antenna, where the distances among array-elements are not strictly required. On the other hand, in order to achieve the maximum SE for massive-MIMO communications, it is required that the distances among antennas in massive-MIMO system are larger than half of the carrier wavelength to achieve the optimal multiplexing of massive-MIMO communications.

This work was supported in part by the National Natural Science Foundation of China (No. 61401330) and the 111 Project of China (B08038).

The authors are with the State Key Laboratory of Integrated Services Networks, Xidian University, Xi'an, 710071, China (corresponding e-mail: wccheng@xidian.edu.cn).

According to different displacement requirements for optimal OAM and massive-MIMO communications, it is clear that OAM and massive-MIMO cannot be entirely equally treated. Moreover, OAM and massive-MIMO are not conflicting with each other [21]. Thus, a question raised is that can we obtain the SE-gain of OAM while earning the SE-gain offered by the traditional multiplexing massive-MIMO communications?

In this paper, we solve the above-mentioned problem and give the answer YES. We build up the OAM embedded massive-MIMO (OEM) communication model, where the MIMO communications are performed among different uniform circular array (UCA) antennas and the OAM communications are performed among the array-elements corresponding to each UCA. Based on the OEM communication model, we design the parabolic antenna to converge the OAM signal. Then, we develop the mode-decomposition and multiplexing-detection scheme to obtain the signal corresponding to each OAM-mode of each transmit UCA antenna. We also develop the OEM-water-filling power allocation policy to obtain the maximum average multiplicative SE-gain for joint OAM and multiplexing massive-MIMO communications. We conduct extensive simulations to validate and evaluate our developed schemes, showing that our developed convergent OEM communications can significantly increase the average SE as compared with traditional multiplexing massive-MIMO communications.

The rest of this paper is organized as follows. Section II gives the OEM communication model. Section III designs the parabolic antenna to converge the OAM wave, proposes the mode-decomposition and multiplexing-detection scheme to obtain the signal on each OAM-mode of each transmit UCA antenna, develops the OEM-water-filling power allocation policy to maximize the average SE of OEM communications, and answers the question when the SE of OEM communication is larger than the SE of traditional massive-MIMO communication. Section IV evaluates our developed schemes and compares the SEs of OEM and traditional multiplexing massive-MIMO communications. The paper concludes with Section V.

## II. THE OAM-EMBEDDED-MIMO COMMUNICATION MODEL

The OEM communication model is shown in Fig. 1, where $N$ and $M$ UCAs [22] are equidistantly around perimeter of the OEM transmit circle and the OEM receive circle, respectively. We denote by $r_1$ the radius of OEM transmit circle and OEM receive circle, $r_2$ the radius of UCAs, $\varphi$ the included angle between the normal line of the OEM transmit circle and the line from the center of the OEM transmit circle to the center of the OEM receive circle, and $\theta$ the included angle between $x$-axis and the projection of the line from the center of the OEM transmit circle to the center of the OEM receive circle on the plane spanned by $x$-axis and $y$-axis. Each UCA at the transmitter is equipped with $U$ array-elements, which are fed with the same input signal, but with a successive delay from array-element to array-element such that after a full turn the phase has been incremented by an integer multiple $u$ of $2\pi$, where $1 \leq u \leq U$. Each UCA at the receiver is equipped with $V$ array-elements with index from $v = 1$ to $v = V$. All array-elements within one transmit or receive UCA antenna share the same RF-chain. Thus, equipped with multiple array-elements, one transmit or receive UCA antenna is actually a single-antenna [20]. The $N$ UCAs at the transmitter and the $M$ UCAs at the receiver can be treated as $N$ transmit and $M$ receive antennas, forming a $N$-transmit-$M$-receive MIMO communication model. All UCAs embed in the MIMO, which is the OEM communication model.

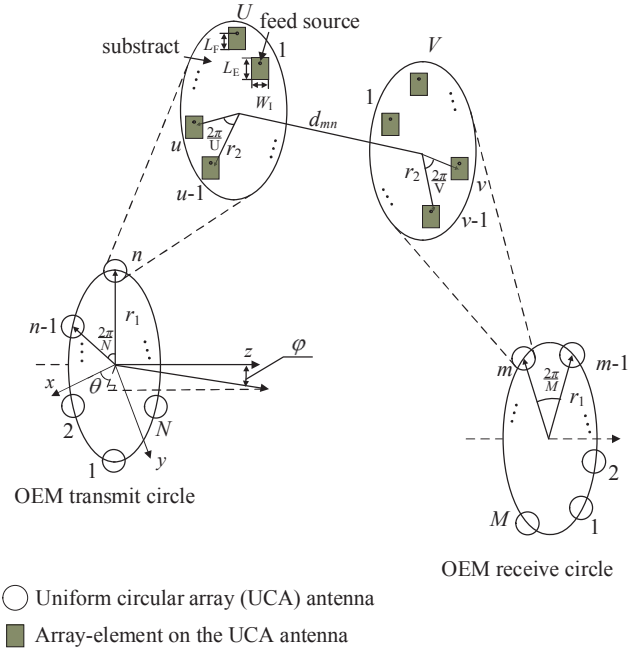

Fig. 1. The OAM-embedded-MIMO communication model.

For each UCA antenna, we need to calculate the width of array-elements, the length of array-elements, and the location of feed source for array-elements. We denote by $W_\mathrm{I}$ the width of array-elements, $L_\mathrm{E}$ the length of array-elements, and $\varepsilon_\mathrm{r}$ the relative permittivity of the substrate, respectively. Then, we can calculate the width of array-elements as follows [23]–[25]:

$$W_\mathrm{I} = \frac{\lambda}{2}\left(\frac{\varepsilon_\mathrm{r}+1}{2}\right)^{-\frac{1}{2}}. \qquad (1)$$

where $\lambda$ denotes the wavelength of carrier. Then, we can calculate the effective permittivity, denoted by $\varepsilon_e$, for the substrate as follows:

$$\varepsilon_\mathrm{e} = \frac{\varepsilon_\mathrm{r}+1}{2} + \frac{\varepsilon_\mathrm{r}-1}{2}\left(1+12\frac{T_\mathrm{H}}{W_\mathrm{I}}\right)^{-\frac{1}{2}}, \qquad (2)$$

where $T_\mathrm{H}$ is the thickness of the substrate and often be set as fixed value. Then, the length of wave guide, denoted by $\lambda_\mathrm{e}$, and $L_\mathrm{E}$, can be obtained from

$$\lambda_\mathrm{e} = \frac{\lambda}{\sqrt{\varepsilon_\mathrm{e}}} \qquad (3)$$

and

$$L_\mathrm{E} = \frac{\lambda_\mathrm{e}}{2} - 2\Delta L_\mathrm{E}, \qquad (4)$$

respectively, where $\Delta L_\text{E}$ is the equivalent length of radiation gap and can be calculated as follows:

$$\Delta L_\text{E} = 0.412 T_\text{H} \frac{(\varepsilon_e + 0.3)(W_\text{I}/T_\text{H} + 0.264)}{(\varepsilon_e - 0.258)(W_\text{I}/T_\text{H} + 0.8)}. \quad (5)$$

The feed source is located $L_\text{F}$ long away from the center of each array-element while the line connecting the center of array-element and the feed source is parallel to the length of the array-element. We denote by $Z_\text{w}$ the wall admittance of array-element and we can calculate $Z_\text{w}$ as follows [23]:

$$Z_\text{w} = \frac{1}{0.00836 \frac{W_\text{I}}{\lambda} + j0.01668 \frac{\Delta L_\text{E} W_\text{I} \varepsilon_e}{T_\text{H} \lambda}}. \quad (6)$$

Then, we define $Z_0$ the characteristic impedance of each array-element. Thus, the input admittance at the feed point, denoted by $Y_1$, can be obtained as follows:

$$Y_1 = \frac{1}{Z_0} \left( \frac{Z_0 \cos \psi L_\text{F} + jZ_\text{w} \sin \psi L_\text{F}}{Z_\text{w} \cos \psi L_\text{F} + jZ_0 \sin \psi L_\text{F}} \right.$$
$$\left. + \frac{Z_0 \cos \psi(L_\text{E} - L_\text{F}) + jZ_\text{w} \sin \psi(L_\text{E} - L_\text{F})}{Z_\text{w} \cos \psi(L_\text{E} - L_\text{F}) + jZ_0 \sin \psi(L_\text{E} - L_\text{F})} \right), \quad (7)$$

where $\psi$ represents the phase constant in medium. Then, the input impedance, denoted by $Z_\text{in}$, for each array-element can be obtained as follows:

$$Z_\text{in} = \frac{1}{Y_1} + j \frac{377}{\sqrt{\varepsilon_\text{r}}} \tan \left( \frac{2\pi T_\text{H}}{\lambda} \right). \quad (8)$$

As a result, when the input impedance for each array-element is equal to 50 Ω, we can obtain $L_\text{F}$ as follows:

$$L_\text{F} = \frac{L_\text{E}}{2} \left(1 - \frac{1}{\sqrt{\xi_\text{re}}}\right), \quad (9)$$

where

$$\xi_\text{re} = \frac{\varepsilon_\text{r} + 1}{2} + \frac{\varepsilon_\text{r} - 1}{2} \left(1 + 12 \frac{T_\text{H}}{L_\text{E}}\right)^{-\frac{1}{2}}. \quad (10)$$

## III. ACHIEVING MAXIMUM MULTIPLICATIVE SPECTRUM EFFICIENCY FOR OEM COMMUNICATIONS

In the section, we mainly propose the mode-decomposition and multiplexing-detection scheme as well as the optimal power allocation policy to maximize the multiplicative SE for convergent OEM communications. First, we design the parabolic antenna based converging method for OAM wave to enable the high-order OAM-mode. Second, we propose the mode-decomposition and multiplexing-detection scheme to resolve the transmit signal on each OAM-mode of each transmit antenna. Third, we develop the OEM-water-filling power allocation policy to maximize the multiplicative SE for OEM communications. At last, we answer the question when do we use the OEM communications.

### A. The Parabolic Antenna Based OAM Wave Converging

According to the principle of OAM communications [10], [26], [27], we give the signal at the $u$th ($1 \leq u \leq U$) array-element on the $n$th ($1 \leq n \leq N$) transmit UCA antenna, denoted by $x_{n,u}$, as follows:

$$x_{n,u} = \sum_{l=0}^{U-1} \frac{1}{\sqrt{U}} s_{n,l} e^{j\phi_u l} = \sum_{l=0}^{U-1} \frac{1}{\sqrt{U}} s_{n,l} e^{j\frac{2\pi(u-1)}{U}l}, \quad (11)$$

where $0 \leq l \leq U - 1$ is the OAM-mode number, $\phi_u$ is the azimuthal angle (defined as the angular position on a plane perpendicular to the axis of propagation) corresponding to the $u$th array-element, and $s_{n,l}$ is the signal on the $l$th OAM-mode of the $n$th transmit UCA antenna. The emitted signal, denoted by $\widetilde{x}_{n,l}$, of the $l$th mode from the $n$th transmit UCA antenna can be treated as a continuous signal expressed as follows:

$$\widetilde{x}_{n,l} = s_{n,l} e^{j\phi l}, \quad (12)$$

where $0 \leq \phi < 2\pi$ is the continuous phase parameter.

For the $n$th transmit UCA antenna and the $m$th receive UCA antenna, we can derive the received signal, denoted by $r_{mn,v,l}$, for the $l$th OAM-mode on the $v$th receive array-element as follows:

$$\begin{aligned} r_{mn,v,l} &= \sum_{u=1}^{U} h_{mn,vu} \frac{1}{\sqrt{U}} s_{n,l} e^{j\frac{2\pi(u-1)}{U}l} \\ &= s_{n,l} \sum_{u=1}^{U} \frac{1}{\sqrt{U}} h_{vu} e^{j\frac{2\pi(u-1)}{U}l} \end{aligned} \quad (13)$$

where $h_{mn,vu}$ denotes the channel amplitude gain corresponding to the $l$th OAM-mode from the $u$th transmit array-element to the $v$th receive array-element. The expression for $h_{mn,vu}$ is given as follows:

$$h_{mn,vu} = \frac{\beta \lambda e^{-j\frac{2\pi}{\lambda}\left|\vec{d}_{mn} - \vec{r}_u + \vec{g}_v\right|}}{4\pi\sqrt{U}\left|\vec{d}_{mn} - \vec{r}_u + \vec{g}_v\right|} \quad (14)$$

where $\beta$ denotes all relevant constants such as attenuation and phase rotation caused by antennas and their patterns on both sides, $\vec{r}_u$ and $\vec{g}_v$ are the vectors from the center of transmit UCA antenna to the $u$th transmit array-element and from the center of receive UCA antenna to the $v$th receive array-element, respectively, where $\vec{d}_{mn}$ is the vector from the center of the $m$th transmit UCA antenna and the center of the $n$th receive UCA antenna, $\left|\vec{d}_{mn} - \vec{r}_u + \vec{g}_v\right|$ represents the distance between the $u$th transmit array-element on the $n$th transmit UCA antenna and the $v$th receive array-element on the $m$th receive UCA antenna. Because of $\left|\vec{d}_{mn}\right| \gg \left|\vec{r}_u\right|$ and $\left|\vec{d}_{mn}\right| \gg \left|\vec{g}_v\right|$, we approximate $\left|\vec{d}_{mn} - \vec{r}_u + \vec{g}_v\right|$ at the denominator of Eq. (14) to $\left|\vec{d}_{mn}\right|$ and $\left|\vec{d}_{mn} - \vec{r}_u + \vec{g}_v\right|$ at the nominator of Eq. (14) to $\left|\vec{d}_{mn}\right| - \vec{d}_{mn}\vec{r}_u / \left|\vec{d}_{mn}\right|$. We denote by $d_{mn} = \left|\vec{d}_{mn}\right|$. Thus, we can rewrite $h_{mn,vu}$ as follows:

$$h_{mn,vu} \approx \frac{\beta \lambda}{4\pi\sqrt{U} d_{mn}} \exp\left[-j\frac{2\pi}{\lambda}\left(d_{mn} - \frac{\vec{d}_{mn}\vec{r}_u}{d_{mn}}\right)\right]. \quad (15)$$

Based on Eqs. (14) and (15), we can derive the equivalent channel amplitude gain, denoted by $h_{mn,llvu,l}$, for the $l$th

OAM-mode corresponding to the $u$th transmit array-element and the $v$th receive array-element as follows:

$$\begin{aligned}
&h_{mn,vu,l} \\
&= \sum_{u=1}^{U} \frac{1}{\sqrt{U}} h_{vu} e^{j\frac{2\pi(u-1)}{U}l} \\
&\approx \frac{\beta\lambda}{4\pi} \sum_{u=1}^{U} \frac{1}{\sqrt{U}} e^{j\frac{2\pi(u-1)}{U}l} \frac{e^{-j\frac{2\pi}{\lambda}d_{mn}} e^{j\frac{2\pi \overrightarrow{d}_{mn}\overrightarrow{r}_u}{\lambda d_{mn}}}}{d_{mn}} \\
&= \frac{\beta\lambda e^{-j\frac{2\pi}{\lambda}d_{mn}}}{4\pi\sqrt{U}d_{mn}} \sum_{u=1}^{U} e^{j\frac{2\pi(u-1)}{U}l} e^{j\frac{2\pi}{\lambda}r_2 \cos\left[\frac{2\pi(u-1)}{U}-\theta\right]\sin\varphi}.
\end{aligned} \tag{16}$$

When $U$ is relatively large enough, we denote by $\theta' = 2\pi(u-1)/U - \theta$ and we have

$$\begin{aligned}
\frac{e^{-j\theta l}j^l}{U} \sum_{u=1}^{U} e^{j\frac{2\pi(u-1)}{U}l} e^{j\frac{2\pi}{\lambda}r_2 \cos\left[\frac{2\pi(u-1)}{U}-\theta\right]\sin\varphi} \\
\approx \frac{j^l}{2\pi} \int_0^{2\pi} e^{j\theta' l} e^{j\frac{2\pi}{\lambda}r_2 \sin\varphi \cos\theta'} d\theta' \\
= J_l\left(\frac{2\pi}{\lambda}r_2 \sin\varphi\right),
\end{aligned} \tag{17}$$

where

$$J_l(\alpha) = \frac{j^l}{2\pi} \int_0^{2\pi} e^{jl\tau} e^{j\alpha\cos\tau} d\tau \tag{18}$$

is the $l$-order Bessel function [28]. Then, based on Eq. (17), we can rewrite $h_{mn,vu,l}$ as follows:

$$h_{mn,vu,l} \approx \frac{\beta\lambda e^{-j\frac{2\pi}{\lambda}d_{mn}}\sqrt{U}e^{j\theta l}j^{-l}}{4\pi d_{mn}} J_l\left(\frac{2\pi}{\lambda}r_2 \sin\varphi\right). \tag{19}$$

Equation (19) shows that the channel amplitude gains corresponding to the $l$th OAM-mode from the $u$th array-element on the $n$th transmit UCA antenna to the $v$th array-element on the $m$th receive UCA antenna are the same for $1 \leq u \leq U$ and $1 \leq v \leq V$. Thus, the UCA based OAM signal can be treated as sending from the center of the transmit UCA antenna and receiving at the center of the receive UCA antenna. Then, we have the channel amplitude gain, denoted by $h'_{mn,l}$, for the $l$th mode corresponding to the $n$th transmit UCA antenna and the $m$th receive UCA antenna as follows:

$$h'_{mn,l} = \frac{\beta\lambda e^{-j\frac{2\pi}{\lambda}d_{mn}}\sqrt{U}e^{j\theta l}j^{-l}}{4\pi d_{mn}} J_l\left(\frac{2\pi}{\lambda}r_2 \sin\varphi\right). \tag{20}$$

Observing Eq. (20), we can find that the transmit UCA antenna causes the OAM signal like going through the Bessel-form channel. According to the characteristics of Bessel function, $h'_{mn,l}$ severely decreases as $l$ increases, resulting in very small received SNR when using the high order (The OAM-mode number is relatively large) OAM communications. That is not what we expected since we aim to obtain the maximum SE combined from all orthogonal OAM-modes. In fact, it has been shown that the electromagnetic wave with OAM is centrally hollow and divergent [29]. Also, as the OAM-mode number increases, the corresponding electromagnetic wave becomes even more divergent. Therefore, if the OAM-mode number is relatively large, it is impossible to get high SE for the OAM communications since the received SNR corresponding to the large OAM-mode number is very small as compared with the PE wave based communications.

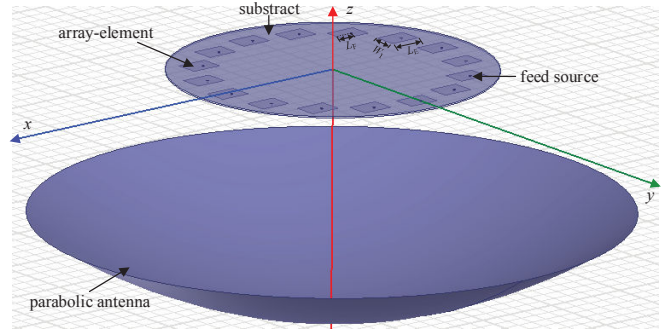

Fig. 2. The designed parabolic antenna for converging OAM waves.

To significantly increase the received SNR for each mode of OAM communications, the electromagnetic waves corresponding to all OAM-modes need to be convergent. Generally, the electromagnetic waves can be converged using the backscatter of parabolic antenna, the refraction of lens antenna, and the diffraction-free super-surface material [30], etc. These schemes can be treated as making a pre-distortion before transmission. Here, we propose the parabolic antenna based converging method to get high SE for high order OAM ($l$ is relatively large) communications. Each UCA needs to equip one parabolic antenna to converge the OAM signal. The center of UCA is placed at the focus of the parabolic antenna with array-elements facing the parabolic antenna. One example for joint UCA and parabolic antenna model is shown in Fig. 2, which is a profile for joint transmit UCA antenna and parabolic antenna model designed in HFSS.

To design the parabolic antenna [31], [32], we denote by $F$ and $D$ the focal length and the diameter of aperture, respectively. Given the gain, denoted by $G$, of parabolic antenna and the efficiency of aperture's utilization, denoted by $\eta$, we can derive the diameter of aperture as follows:

$$D = \frac{\lambda\sqrt{G\eta}}{\pi}. \tag{21}$$

Then, we can obtain the focal length as follows:

$$F = \kappa D, \tag{22}$$

where $\kappa$ is the focal length to aperture diameter ratio of parabolic antenna. The value of $\kappa$ is often chosen between 0.25 and 0.5. Then, the equation for parabolic antenna can be derived as follows:

$$Z = (X^2 + Y^2)/4F, \tag{23}$$

where $X$, $Y$, and $Z$ are the $x$-axis, $y$-axis, and $z$-axis coordinates for the parabolic antenna.

After converging by the parabolic antenna, the impact of item $J_l(2\pi r_2 \sin\varphi/\lambda)$ can be greatly reduced since the value of angle $\varphi$ is greatly reduced. We denote by $\varphi_c$ the equivalent angle for convergent OEM communications. Then, the equivalent channel amplitude gain, denoted by $\widetilde{h}_{mn,l}$, for the $l$th mode corresponding to the $n$th transmit UCA antenna and the $m$th receive UCA antenna can be modeled as follows:

$$\widetilde{h}_{mn,l} = \frac{A_l \beta \lambda e^{-j\frac{2\pi}{\lambda} d_{mn}} \sqrt{U} e^{j\theta l} j^{-l}}{4\pi d_{mn}} J_l\left(\frac{2\pi}{\lambda} r_2 \sin\varphi_c\right), \quad (24)$$

where $A_l$ contains the amplitude gain on the $l$th OAM-mode caused by the backscatter of parabolic antenna.

### B. The Mode-Decomposition and Multiplexing-Detection Scheme

At the receiver, the convergent vorticose signal can be spatially sampled at each receive UCA antenna. Then, after the sampling, the signal received at the $v$th array-element of the $m$th receive UCA antenna, denoted by $y_{m,v}$, can be derived as follows:

$$y_{m,v} = \sum_{l=0}^{U-1} \sum_{n=1}^{N} \widetilde{h}_{mn,l} s_{n,l} e^{j\frac{2\pi(v-1)}{V}l} + w_{m,v} \quad (25)$$

where $w_{m,v}$ denotes the received noise at the $v$th array-element of the $m$th receive UCA antenna. To obtain the received signal on the $l_0$th mode ($0 \leq l_0 \leq U-1$) sent from all transmit UCA antennas, we multiply $y_{m,v}$ with $\exp\{-j[2\pi(v-1)l_0]/V\}$ and we have the received signal, denoted by $y_{m,v,l_0}$, on the $l_0$ mode corresponding to the $v$th array-element of the $m$th received UCA as follows:

$$\begin{aligned} y_{m,v,l_0} &= y_{m,v} e^{-j\frac{2\pi(v-1)}{V}l_0} \\ &= \sum_{n=1}^{N} \sum_{l=0, l\neq l_0}^{U-1} \widetilde{h}_{mn,l} s_{n,l} e^{j\frac{2\pi(v-1)}{V}(l-l_0)} \\ &\quad + \sum_{n=1}^{N} \widetilde{h}_{mn,l_0} s_{n,l_0} + w_{m,v} e^{-j\frac{2\pi(v-1)}{V}l_0}. \end{aligned} \quad (26)$$

Then, the received signal, denoted by $\widetilde{y}_{m,l_0}$, on the $l_0$th mode of the $m$th receive UCA antenna can be derived as follows:

$$\widetilde{y}_{m,l_0} = \sum_{v=1}^{V} y_{m,v,l_0} = \sum_{v=1}^{V} \sum_{n=1}^{N} \widetilde{h}_{mn,l_0} s_{n,l_0} + \widetilde{w}_{m,l_o}, \quad (27)$$

where $\widetilde{w}_{m,l_0}$ denotes the received noise on the $l_0$th mode corresponding to the $m$th receive UCA antenna. Now, we have obtained the estimated decomposed signal, specified in Eq. (27), for all OAM-modes ($0 \leq l_0 \leq U-1$). It is clear that $\widetilde{y}_{m,l_0}$ follows the standard form of received signal as that in multiplexing-MIMO communications.

We denote by $\boldsymbol{y}_l = [\widetilde{y}_{1,l}, \widetilde{y}_{2,l}, \cdots, \widetilde{y}_{M,l}]^T$ and $\boldsymbol{w}_l = [\widetilde{w}_{1,l}, \widetilde{w}_{2,l}, \cdots, \widetilde{w}_{M,l}]^T$ the received signal and noise, respectively, corresponding to the $l$th mode, where $(\cdot)^T$ represents the transpose operation. Using the zero-forcing detection scheme, the estimate value of transmit signal, denoted by $\widehat{\boldsymbol{s}_l}$, can be derived as follows:

$$\begin{aligned} \widehat{\boldsymbol{s}_l} &= (\boldsymbol{H}_l^H \boldsymbol{H}_l)^{-1} \boldsymbol{H}_l^H \boldsymbol{y}_l \\ &= \boldsymbol{s}_l + (\boldsymbol{H}_l^H \boldsymbol{H}_l)^{-1} \boldsymbol{H}_l^H \boldsymbol{w}_l \end{aligned} \quad (28)$$

where $(\cdot)^H$ denotes the conjugation operation, $\boldsymbol{w}_l$ is the received noise on the $l$th mode, and

$$\boldsymbol{H}_l = V \begin{bmatrix} \widetilde{h}_{11,l} & \widetilde{h}_{12,l} & \cdots & \widetilde{h}_{1N,l} \\ \widetilde{h}_{21,l} & \widetilde{h}_{22,l} & \cdots & \widetilde{h}_{2N,l} \\ \vdots & \vdots & \ddots & \vdots \\ \widetilde{h}_{M1,l} & \widetilde{h}_{M2,l} & \cdots & \widetilde{h}_{MN,l} \end{bmatrix} \quad (29)$$

is the amplitude gain matrix of all channels on the $l$th mode. Then, the received channel signal-to-noise-ratio (SNR), denoted by $\text{SNR}_l$, for the $l$th model can be derived as follows:

$$\text{SNR}_l = \frac{|\boldsymbol{s}_l|^2}{\sigma_l^2 |(\boldsymbol{H}_l^H \boldsymbol{H}_l)^{-1} \boldsymbol{H}_l^H|^2}, \quad (30)$$

where $\sigma_l^2$ denotes the variance of received noise corresponding to the $l$th mode.

### C. The OEM-Water-Filling Power Allocation Policy

For the OEM communications, the instantaneous SE, denoted by $C_{\text{OEM}}$, can be derived as follows:

$$\begin{aligned} C_{\text{OEM}} &= \sum_{l=0}^{U-1} \log_2 \det\left(I + \frac{|\boldsymbol{s}_l|^2}{\sigma_l^2 |(\boldsymbol{H}_l^H \boldsymbol{H}_l)^{-1} \boldsymbol{H}_l^H|^2}\right) \\ &= \sum_{l=0}^{U-1} \sum_{i=1}^{\text{rank}(\boldsymbol{H}_l)} \log_2\left(1 + P_{i,l}\gamma_{i,l}\right), \end{aligned} \quad (31)$$

where $I$ is the identity matrix, $\gamma_{i,l} = \overline{P}/[\sigma_l^2 |(\boldsymbol{H}_l^H \boldsymbol{H}_l)^{-1} \boldsymbol{H}_l^H|^2]$ represents the received SNR corresponding to matrix $\boldsymbol{H}_l$ with $\overline{P}$ representing the upper-bound of total power, $\text{rank}(\boldsymbol{H}_l)$ denotes the rank of $\boldsymbol{H}_l$, and $P_{i,l}$ denotes the power allocation policy. We aim to maximize the SE for OEM communications. Thus, we formulate the average SE maximization problem, denoted by **P1**, as follows:

$$\textbf{P1:} \quad \max \mathbb{E}_{\boldsymbol{\gamma}}\left[\sum_{l=0}^{U-1} \sum_{i=1}^{\text{rank}(\boldsymbol{H}_l)} \log_2\left(1 + P_{i,l}\gamma_{i,l}\right)\right] \quad (32)$$

$$\text{s.t.} : 1). \mathbb{E}_{\boldsymbol{\gamma}}\left[\sum_{l=0}^{U-1} \sum_{i=1}^{\text{rank}(\boldsymbol{H}_l)} P_{i,l}\right] \leq \overline{P}; \quad (33)$$

$$2). P_{i,l} \geq 0, \forall l \in [0, U-1], \forall i \in [1, \text{rank}(\boldsymbol{H}_l)], \quad (34)$$

where

$$\boldsymbol{\gamma} = \begin{bmatrix} \gamma_{1,0} & \gamma_{1,1} & \cdots & \gamma_{1,U-1} \\ \gamma_{2,0} & \gamma_{2,1} & \cdots & \gamma_{2,U-1} \\ \vdots & \vdots & \ddots & \vdots \\ \gamma_{\text{rank}(\boldsymbol{H}_l),0} & \gamma_{\text{rank}(\boldsymbol{H}_l),1} & \cdots & \gamma_{\text{rank}(\boldsymbol{H}_l),U-1} \end{bmatrix} \quad (35)$$

is the instantaneous received SNR for all channels of the OEM communications and $\mathbb{E}_{\boldsymbol{\gamma}}\{\cdot\}$ is the expectation operation with respect to $\boldsymbol{\gamma}$. We assume that the channels corresponding to all modes of all UCAs follow the Rayleigh distribution, where the probability density function (PDF), denoted by $p(\gamma_{i,l})$, is given by $p(\gamma_{i,l}) = \exp(-\gamma_{i,l}/\overline{\gamma}_{i,l})/\overline{\gamma}_{i,l}$ and $\overline{\gamma}_{i,l}$ denotes the average received SNR.

6It is clear that **P1** is a strictly convex optimization problem [33]. To solve **P1**, we construct the Lagrangian function for **P1**, denoted by $J$, as follows:

$$J = \mathbb{E}_{\gamma}\left[\sum_{l=0}^{U-1}\sum_{i=1}^{\text{rank}(\boldsymbol{H}_l)} \log_2\left(1 + P_{i,l}\gamma_{i,l}\right)\right] + \sum_{i=1}^{\text{rank}(\boldsymbol{H}_l)}\sum_{l=0}^{U-1} \epsilon_{i,l} P_{i,l}$$
$$-\mu\left(\mathbb{E}_{\gamma}\left[\sum_{l=0}^{U-1}\sum_{i=1}^{\text{rank}(\boldsymbol{H}_l)} P_{i,l}\right] - \overline{P}\right), \quad (36)$$

where $\mu \geq 0$ and $\epsilon_{i,l} \geq 0$ ($1 \leq i \leq \text{rank}(\boldsymbol{H}_l)$, $0 \leq l \leq U-1$) are the Lagrangian multipliers associated with the constraints specified by Eqs. (33) and (34), respectively. Taking the derivative for $J$ with the respect to $P_{i,l}$ and setting the derivative equal to zero, we can obtain a set of $\sum_{l=0}^{U-1} \text{rank}(\boldsymbol{H}_l)$ equations as follows:

$$\frac{\partial J}{\partial P_{i,l}} = \frac{\frac{1}{\log 2}}{1 + P_{i,l}\gamma_{i,l}}\gamma_{i,l}p_{\Gamma}(\gamma) - \mu p_{\Gamma}(\gamma) + \epsilon_{i,l} = 0, \quad (37)$$

where $p_{\Gamma}(\gamma)$ is the probability density function corresponding to the channel. According to the principle of complementary slackness, we have $\epsilon_{i,l} P_{i,l} = 0$ for $\forall l \in [0, U-1]$ and $\forall i \in [1, \text{rank}(\boldsymbol{H}_l)]$. Correspondingly, we consider two different cases as follows:

**Case A:** The inequality $P_{i,l} > 0$ holds for $\forall l \in [0, U-1]$ and $\forall i \in [1, \text{rank}(\boldsymbol{H}_l)]$. Under this case, all channels corresponding to all OAM-modes of all transmit UCA antennas are assigned non-zero power for data transmission. Thus, based on the complementary slackness, we have $\epsilon_{i,l} = 0$ for $\forall l \in [0, U-1]$ and $\forall i \in [1, \text{rank}(\boldsymbol{H}_l)]$. Then, Eq. (37) can be reduced to

$$\frac{\frac{\gamma_{i,l}}{\log 2}}{1 + P_{i,l}\gamma_{i,l}} - \mu^* = 0. \quad (38)$$

where $\mu^*$ is the optimal Lagrangian multiplier corresponding to $\mu$. Solving Eq. (38), we can obtain the optimal power allocation policy for **Case A** as follows:

$$P_{i,l} = \frac{1}{\mu^* \log 2} - \frac{1}{\gamma_{i,l}},$$
$$\text{for } 1 \leq i \leq \text{rank}(\boldsymbol{H}_l) \text{ and } 0 \leq l \leq U-1, \quad (39)$$

where the optimal value for $\mu^*$ can be numerically obtained by substituting $P_{i,l}$ into

$$\mathbb{E}_{\gamma}\left[\sum_{l=0}^{U-1}\sum_{i=1}^{\text{rank}(\boldsymbol{H}_l)} P_{i,l}\right] = \overline{P}. \quad (40)$$

We define

$$\mathcal{N}_1 = \{(i,l) | 1 \leq i \leq \text{rank}(\boldsymbol{H}_l), 0 \leq l \leq U-1\} \quad (41)$$

and we also define $\mathcal{N}_2$ as the index-set which satisfies the strict inequality as follows:

$$\mathcal{N}_2 \triangleq \left\{(i,l) \in \mathcal{N}_1 \left| \frac{1}{\mu^* \log 2} - \frac{1}{\gamma_{i,l}} > 0 \right.\right\}. \quad (42)$$

Then, Eq. (39) is the optimal solution only if $\mathcal{N}_2 = \mathcal{N}_1$. Otherwise, if $\mathcal{N}_2 \subset \mathcal{N}_1$, we need to consider the following case.

**Case B:** There exist some $P_{i,l}$ such that $P_{i,l} = 0$. If $\mathcal{N}_2 \subset \mathcal{N}_1$, there certainly exist some $P_{i,l}$ such that $P_{i,l} = 0$. That means some OAM-modes (maybe some transmit UCAs) are not assigned power and are not used for data transmission. We give the following lemma.

*Lemma 1:* If $(i,l) \notin \mathcal{N}_2$, then $P_{i,l} = 0$.

*Proof:* We need to show that for any non-empty subset $\mathcal{C} \subseteq \overline{\mathcal{N}_2}$, there is no power allocation policy $P_{i,l}$ such that $P_{i,l} > 0$ for all $(i,l) \in (\mathcal{C} \cup \mathcal{N}_2)$.

If $\mathcal{C} = \overline{\mathcal{N}_2}$, we have already known that there is no power allocation policy.

Otherwise, if $\mathcal{C} \subset \overline{\mathcal{N}_2}$, we suppose that there exists such a power allocation policy so that it can be expressed as

$$P_{i,l} = \begin{cases} \frac{1}{\mu^* \log 2} - \frac{1}{\gamma_{i,l}}, & (i,l) \in (\mathcal{C} \cup \mathcal{N}_2); \\ 0, & \text{otherwise.} \end{cases} \quad (43)$$

According to Eq. (43), we can select an element $(i', l') \in \mathcal{C}$, which satisfies $P_{i',l'} > 0$. Thus, we can obtain

$$\gamma_{i',l'} > \mu^* \log 2. \quad (44)$$

On the other hand, according the definition of $\mathcal{N}_2$ specified in Eq. (42) and $\overline{\mathcal{N}_2}$, we know

$$\gamma_{i',l'} < \mu^* \log 2, \quad (45)$$

which is opposed to Eq. (44). Therefore, such a power allocation policy does not exist and Lemma 1 follows. ∎

Following the same procedure as that used in **Case A**, if the strict inequality $P_{i,l} > 0$ holds for all $(i,l) \in \mathcal{N}_2$, we can obtain the optimal power allocation policy as follows:

$$P_{i,l} = \begin{cases} \frac{1}{\mu^* \log 2} - \frac{1}{\gamma_{i,l}}, & (i,l) \in \mathcal{N}_2; \\ 0, & \text{otherwise.} \end{cases} \quad (46)$$

Otherwise, if not all $(i,l) \in \mathcal{N}_2$ satisfy the strict inequality $P_{i,l} > 0$, we need to further divide $\mathcal{N}_2$ and repeat the procedure again. In summary, the optimal power allocation policy, called the OEM-water-filling power allocation policy, is given by Algorithm 1 in the following.

---

**Algorithm 1** : The OEM-Water-Filling Power Allocation Policy
---
1) **Initialization**
   a) Obtain $\mathcal{N}_2$ by Eq. (42)
   b) $k = 2$
2) **While** ($\mathcal{N}_k \neq \mathcal{N}_{k-1}$)
   a) $\mathcal{N}_{k+1} = \left\{(i,l) \in \mathcal{N}_k \left| \frac{1}{\mu^* \log 2} - \frac{1}{\gamma_{i,l}} > 0 \right.\right\}$
   b) $k = k +1$
3) **Obtain the optimal power allocation policy**
   a) Denote $\mathcal{N}^* = \mathcal{N}_k$
   b)
   $$lP_{i,l} = \begin{cases} \frac{1}{\mu^* \log 2} - \frac{1}{\gamma_{i,l}}, & (i,l) \in \mathcal{N}^*; \\ 0, & \text{otherwise.} \end{cases}$$
---

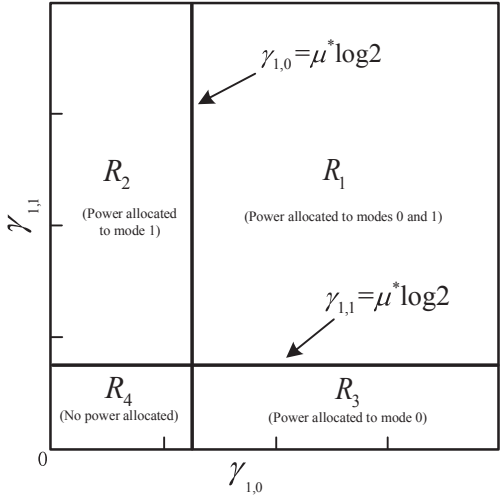

Fig. 3. The regions for the OEM-water-filling power allocation policy corresponding to the two OAM-modes case ($l=0$ and $l=1$).

To show the execution procedure of our developed OEM-water-filling power allocation policy, we demonstrate a particular case when $i=1$ and $l \in \{0,1\}$ for OEM communications. Using Algorithm 1, we can see that the optimal power allocation policy partitions the SNR-plane $(\gamma_{1,0}, \gamma_{1,1})$ into four exclusive regions by the lines as shown in Fig. 3. If $(\gamma_{1,0}, \gamma_{1,1})$ falls into region $R_1$, both two subchannels corresponding to $l=0$ and $l=1$ will be assigned with power for data transmission, where the boundaries of region $R_1$ is determined by $\gamma_{1,0}=\mu^*\log 2$ and $\gamma_{1,1}=\mu^*\log 2$ which are obtained by solving the boundary condition $\mathcal{N}_2 = \mathcal{N}_1$. On the other hand, if $(\gamma_{1,0}, \gamma_{1,1})$ falls into either $R_2$ or $R_3$, then only one of the subchannels will be assigned with power. Otherwise, if $(\gamma_{1,0}, \gamma_{1,1})$ belongs to region $R_4$, there is no power allocated to any subchannel and the system will be in an outage state.

Substituting the OEM-water-filling power allocation policy into Eq. (31), we can obtain the maximum average SE, denoted by $C^*_{\text{OEM}}$, for OEM communications as follows:

$$C^*_{\text{OEM}} = \mathbb{E}_\gamma \left[ \sum_{l=0}^{U-1} \sum_{i=1}^{\text{rank}(\boldsymbol{H}_l)} \log_2 \left( \frac{\gamma_{i,l}}{\mu^* \log 2} \right) \right]. \qquad (47)$$

To compare the SE of OEM communications with that of the traditional multiplexing-MIMO communications, we also give the maximum average SE, denoted by $C^*_{\text{MIMO}}$ for the traditional multiplexing-MIMO communications as follows:

$$C^*_{\text{MIMO}} = \sum_{i=1}^{\text{rank}(\widetilde{\boldsymbol{H}})} \log_2 \left( 1 + \widetilde{P}_i \widetilde{\gamma}_i \right), \qquad (48)$$

where $\text{rank}(\widetilde{\boldsymbol{H}})$ denotes rank corresponding to the MIMO communication channel matrix $\widetilde{\boldsymbol{H}}$, $\widetilde{P}_i$ is the optimal power allocation [34] for MIMO communications and $\widetilde{\gamma}_i$ is the received SNR corresponding to matrix $\widetilde{\boldsymbol{H}}$.

### D. When Do We Use the OEM Communications?

The traditional multiplexing-MIMO requires that the distance between two adjacent antennas needs to be not less than half of the carrier wavelength [35], [36], which limits the displacement of multiple antennas in MIMO systems. However, there is no minimum distance restriction on OAM vorticose communications, thus bridging the new way to achieve very high SE.

As for our proposed OEM model, if the distance between two adjacent UCA array-elements is less than half of the carrier wavelength, the UCA-based OAM vorticose communications can be embedded into the massive-MIMO system, forming the OEM communication to achieve larger SE than the traditional massive-MIMO system.

We denote by $d_a(r_1, N)$ and $d_e(r_2, U)$ the distances between the centers of two adjacent UCAs and between the center of two adjacent array-elements, respectively. Then, we have

$$\begin{cases} d_a(r_1, N) = \sqrt{r_1^2 + r_1^2 - 2r_1^2 \cos \frac{2\pi}{N}}; \\ d_e(r_2, U) = \sqrt{r_2^2 + r_2^2 - 2r_2^2 \cos \frac{2\pi}{U}}. \end{cases} \qquad (49)$$

Based on Eq. (49), we can analyze two scenarios as follows:
**Scenario I:** $d_a(r_1, N) > \lambda/2$ and $d_e(r_2, U) \leq \lambda/2$. Under this scenario, the OAM signal can be generated by every UCA antenna. However, multiple antennas cannot be equipped at the position of UCA due to the half length distance restriction. The OEM communications can significantly increase the SE as compared with the traditional massive-MIMO communications. For fixed $r_1, r_2, U$, and $N$, we can obtain the wavelength region for applying the OEM communications as follows:

$$r_1 \sqrt{8 \left( 1 - \cos \frac{2\pi}{U} \right)} < \lambda < r_2 \sqrt{8 \left( 1 - \cos \frac{2\pi}{N} \right)} \qquad (50)$$

**Scenario II:** $d_a(r_1, N) > \lambda/2$ and $d_e(r_2, U) > \lambda/2$. In this case, we do not need to use OEM communications since the traditional massive-MIMO has achieved the maximum SE.

## IV. PERFORMANCE EVALUATIONS

In this section, we evaluate the performance of our designed parabolic antenna converging and developed OEM-water-filling power allocation policy for achieving the maximum multiplicative SE in OEM communications. Throughout our evaluations, we set the frequency of carriers as 35 GHz and the angle $\varphi$ as $30°$. We use HFSS to evaluate the converging and MATLAB to evaluate the multiplicative SE, respectively. The circular array consists of sixteen identical rectangular patches (array-element), which are excited by coaxial feed method [23]. The UCA is designed using F4B substrate with a relative permittivity of 2.2 and a thickness of 0.245 mm [23]. We also set $G = 36$ dB, $\eta = 50\%$, and $k = 0.4$, respectively.

Figure 4 shows the directional diagrams of different OAM-modes ($l=0,1,2,3$) for non-convergent and convergent OEM waves with $N=1$ and $M=1$, where we set $U=V=16$. As we expected, the maximum gain of non-convergent OEM wave decreases as the OAM-mode number increases.

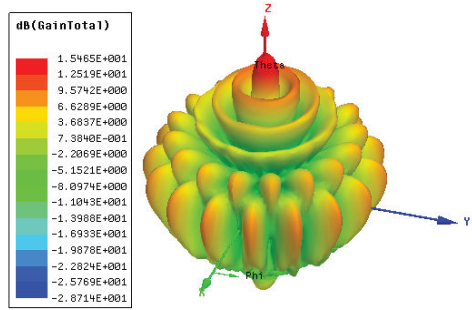

(a) Mode 0 (Non-convergent OEM, $l = 0$)

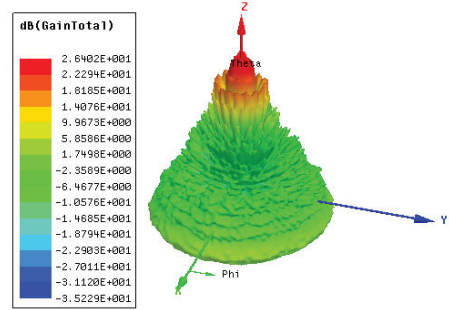

(b) Mode 0 (Convergent OEM, $l = 0$)

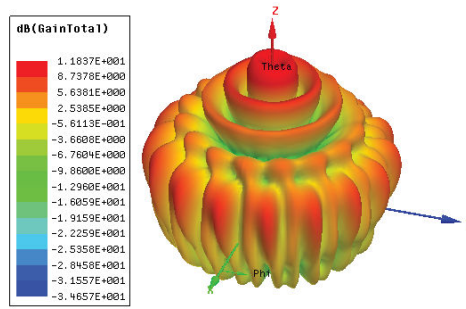

(c) Mode 1 (Non-convergent OEM, $l = 1$)

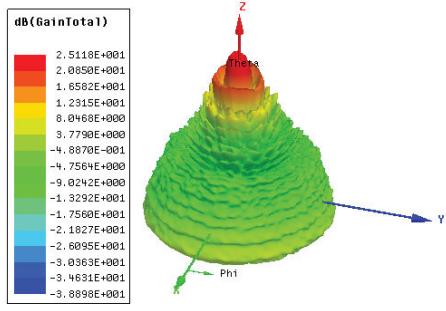

(d) Mode 1 (Convergent OEM, $l = 1$)

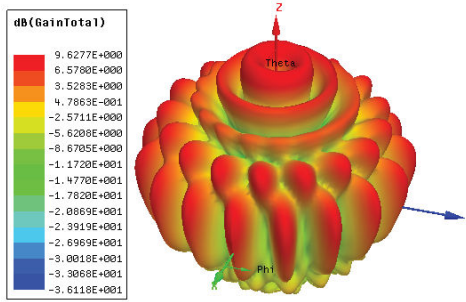

(e) Mode 2 (Non-convergent OEM, $l = 2$)

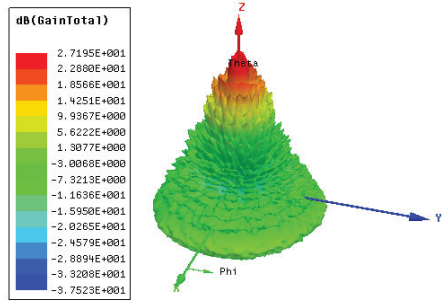

(f) Mode 2 (Convergent OEM, $l = 2$)

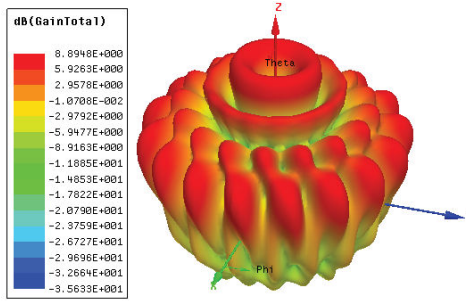

(g) Mode 3 (Non-convergent OEM, $l = 3$)

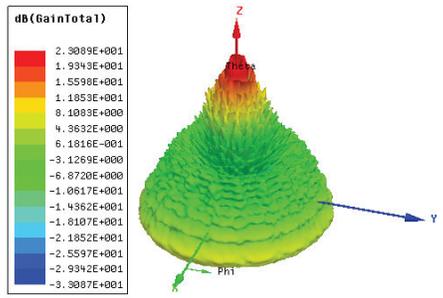

(h) Mode 3 (Convergent OEM, $l = 3$)

Fig. 4. The directional diagram for non-convergent and convergent OEM waves with $N = 1$ and $M = 1$.

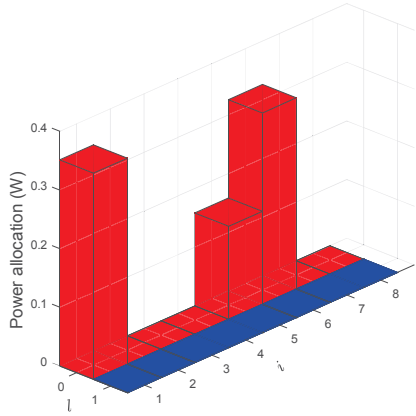

(a) Non-convergent OEM, $N = M = 8(U = V = 2)$

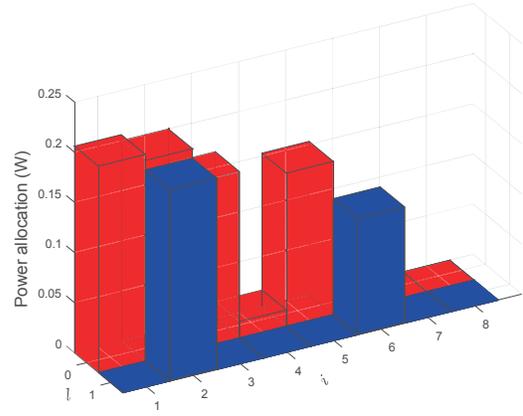

(b) Convergent OEM, $N = M = 8(U = V = 2)$

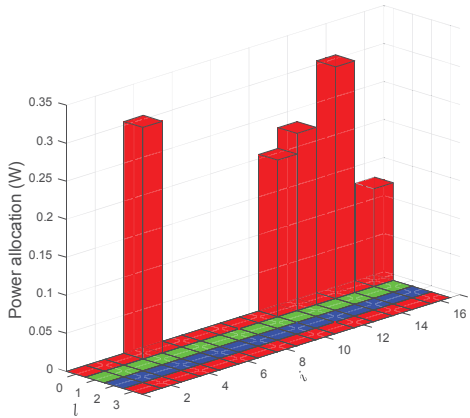

(c) Non-convergent OEM, $N = M = 16(U = V = 4)$

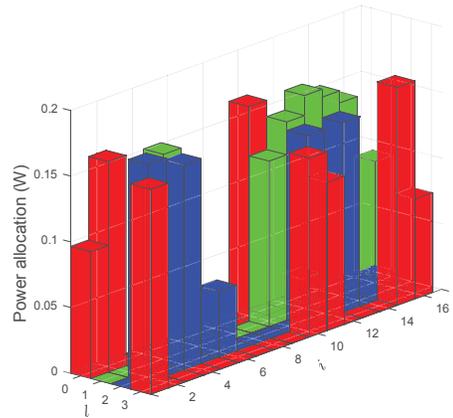

(d) Convergent OEM, $N = M = 16(U = V = 4)$

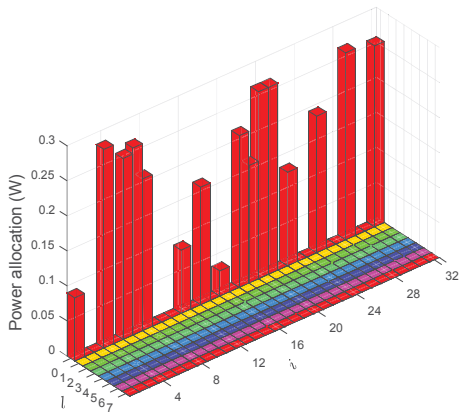

(e) Non-convergent OEM, $N = M = 32(U = V = 8)$

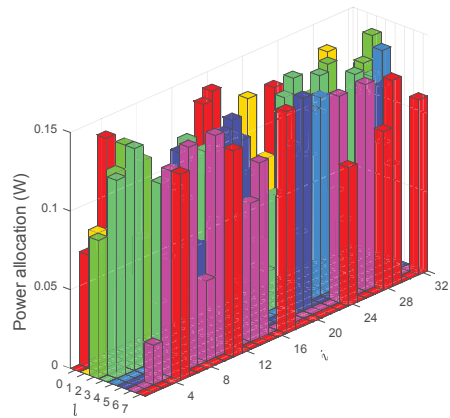

(f) Convergent OEM, $N = M = 32(U = V = 8)$

Fig. 5. The three cases of instantaneous OEM-water-filling power allocations for non-convergent and convergent OEM communications.

The array mainly radiates along the $z$-axis direction without a central hollow when mode number $l = 0$. This is because $l = 0$ OAM wave is the traditional PE wave. The central hollow becomes larger and larger as the order of OAM increases. The radius of central hollow is significantly reduced for the convergent OAM waves and the radiated power concentrates on the main lobe with little power allocated on the side lobe after converging.

Figure 5 depicts our developed OEM-water-filling power allocation policy for given instantaneous channel SNR in OEM communications, where we set $N = M = 8$ ($U = V = 2$), $N = M = 16$ ($U = V = 4$), and $N = M = 32$ ($U = V = 8$), respectively. The average transmit power is given by $\overline{P} = 0.2$ W. As shown in Fig. 5 before converging, the allocated power severely decreases as the order of OAM increases. There is little power allocated to the high order OAM-modes in OEM communications. This is because the average received SNRs on high order of OAM-modes are very small, which is consistent with the property of Bessel functions. After converging, almost all OAM-modes can be allocated part of power. Thus, all OAM-modes can be used to increase the SE for OEM communications.

Figure 6 shows our developed OEM-water-filling power allocation versus the instantaneous received channel SNR, where we set $N = M = 32$ ($U = V = 4$ and $U = V = 2$, respectively) and $N = M = 16$ ($U = V = 4$ and $U = V = 2$, respectively) for OEM communications. As shown in Fig. 6, the power allocation of OEM-water-filling policy increases as the instantaneous channel received SNR increases until reaching to a maximum value and holds on. The allocated power for $U = V = 4$ OEM communications is less than that for corresponding $U = V = 2$ OEM comminations. This is because the power is generally averaged allocated to different OAM-modes. The allocated power for $N = M = 32$ ($U = V = 2$) OEM communications and $N = M = 16$ ($U = V = 4$) OEM communications are the same. This is because both of them have the same number of orthogonal channels.

Figure 7 depicts the average SEs of non-convergent OEM communications and traditional massive-MIMO communications, where we set $U = V = 2$ and $M = N = 8$, 16, 32, and 64, respectively. Observing the curves in Fig. 7, all SEs for non-convergent OEM and traditional massive-MIMO increase as the average channel SNR increases. More importantly, when equipping with the same transmit-UCAs and receive-UCAs, the non-convergent OEM and the traditional massive-MIMO obtain the same SE, as verified for $M = N = 8$, 16, 32, and 64 cases in Fig. 7. The reason why the SEs of non-convergent OEM and traditional massive-MIMO are the same is that much power is allocated to the low order OAM-mode while the high order OAM-mode is allocated almost zero power. Therefore, it is necessary to converge OEM waves to efficiently use the high order OAM-mode to achieve maximum SE for OEM communications.

Figure 8 compares the average SEs of convergent OEM and traditional massive-MIMO communications, where we set $N = M = 32$ ($U = V = 8$ and $U = V = 4$, respectively) for OEM communication, $N = M = 16$ ($U = V = 4$) for OEM communication, and $N = M = 32$ for massive-MIMO communication, respectively. As shown in Fig. 8, the OEM communications significantly increase average SEs as compared with the traditional massive-MIMO communications. The $N = M = 16$ ($U = V = 4$), $N = M = 32$ ($U = V = 4$), and $N = M = 32$ ($U = V = 8$) OEM communications achieve double, 4 times, 8 times SEs, respectively, as compared with the $N = M = 32$ massive-MIMO communication. This is because the OEM communication can obtain not only the SE of traditional massive-MIMO communication, but also the SE of embedded OAM communication. The joint OAM and MIMO communications offers the multiplicative SE which is the SE of traditional massive-MIMO communication times the SE of embedded OAM communication.

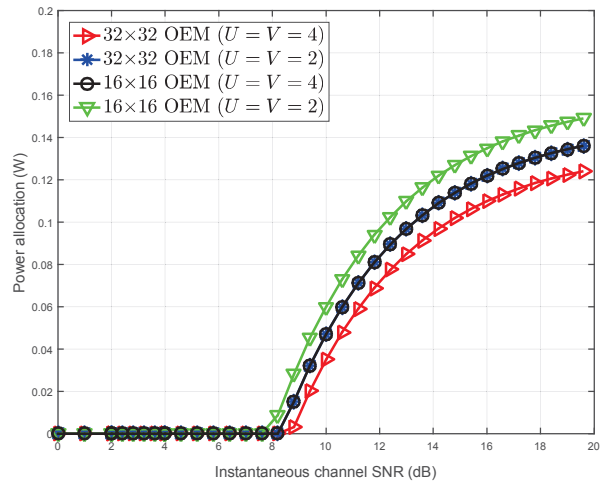

Fig. 6. The OEM-water-filling power allocation versus instantaneous channel SNR.

## V. CONCLUSIONS

We proposed the framework for OEM communications to obtain multiplicative spectrum-efficiency gains for joint OAM and massive-MIMO mmWave wireless communications. We designed the parabolic antenna to converge the OAM waves. Then, we developed the joint mode-decomposition and multiplexing-detection scheme to decompose and detect the transmit signal on each OAM-mode of each transmit UCA antenna. We also developed the OEM-water-filling power allocation policy to achieve the maximum multiplicative spectrum-efficiency for OEM communications. The simulation results validated the parabolic antenna based converging, the joint mode-decomposition and multiplexing-detection scheme, and the OEM-water-filling policy. The OEM communications can obtain the multiplicative spectrum-efficiency gain for joint OAM and massive-MIMO communications, thus achieving much larger spectrum-efficiency than the traditional massive-MIMO mmWave communications.

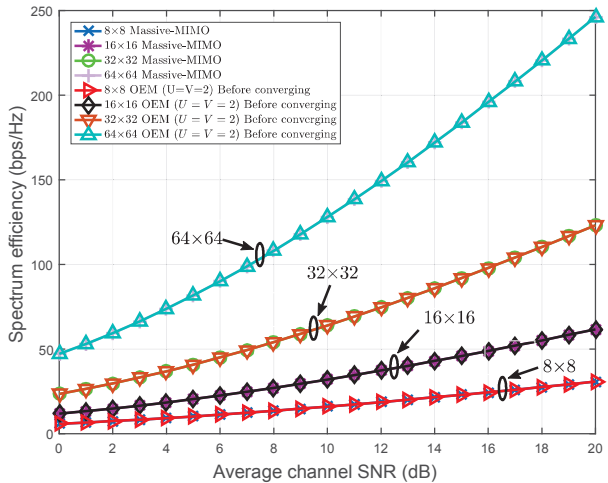

Fig. 7. The average spectrum-efficiencies of non-convergent OEM and traditional massive-MIMO.

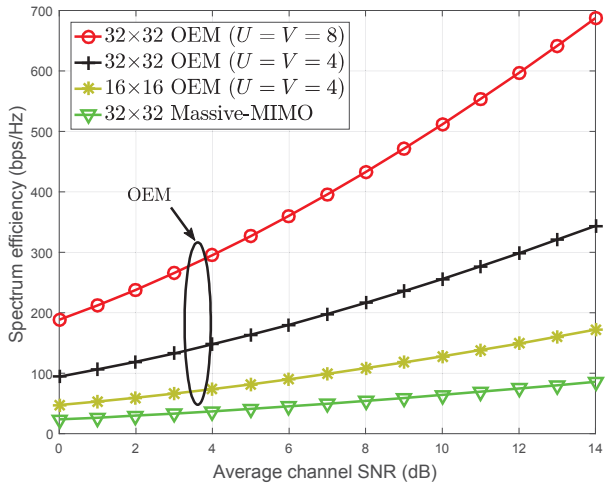

Fig. 8. The average spectrum-efficiency versus the average channel SNR.